\documentclass[submission,copyright,creativecommons]{eptcs}
\providecommand{\event}{AREA 2023} 

\usepackage{iftex}

\ifpdf
  \usepackage{underscore}         
  \usepackage[T1]{fontenc}        
\else
  \usepackage{breakurl}           
\fi

\newtheorem{definition}{Definition}
\newcommand{\EATL}[1]{{\langle\!\langle #1 \rangle\!\rangle}}
\newcommand{\AATL}[1]{{[\![ #1 ]\!]}}
\newcommand{\ESL}[1]{\exists\:\! {#1}}
\newcommand{\ASL}[1]{\forall\:\!\!{#1}}

\title{The Impact of Strategies and Information \\ in Model Checking for Multi-Agent Systems}
\author{Vadim Malvone
\institute{LTCI, Télécom Paris, Institut Polytechnique de Paris, \\ Palaiseau, France}
\email{vadim.malvone@telecom-paris.fr}
}
\def\titlerunning{The Impact of Strategies and Information in Model Checking for MAS}
\def\authorrunning{Vadim Malvone}
\begin{document}
\maketitle

\begin{abstract}
System correctness is one of the most crucial and challenging objectives in software and hardware systems. With the increasing evolution of connected and distributed systems, ensuring their correctness requires the use of formal verification for multi-agent systems.
In this paper, we present a summary of certain results on model checking for multi-agent systems that derive from the selection of strategies and information for agents. Additionally, we discuss some open directions for future research.
\end{abstract}

\section{Introduction}
The problem of assuring systems correctness is particularly felt in hardware and software design, especially in safety-critical scenarios. When we talk about a safety-critical system, we mean the one in which failure is not an option. To face this problem, several methodologies have been proposed. Among these, Model Checking (MC)~\cite{clarke1999model,clarke2018model} results to be very useful. This approach provides a formal-based methodology to model systems, to specify properties via temporal logics, and to verify that a system satisfies a given specification.

Notably, first applications of model checking just concerned closed systems, which are characterized by the fact that their behavior is completely determined by their internal states. Unfortunately, model checking techniques developed to handle closed systems turn out to be quite useless in practice, as most of the systems are open and are characterized by an ongoing interaction with other systems. To overcome this problem, model checking has been extended to Multi-Agent Systems (MAS). In the latter context, temporal logics have been extended to temporal logics for the strategic reasoning such as Alternating-time Temporal Logic (ATL)~\cite{AlurHK02} and Strategy Logic (SL)~\cite{MogaveroMPV14}. 

Given the logics under exam, there are two key aspects in MAS to determine the model checking complexity: the type of strategies and the agents' information.
A strategy is a generic conditional plan that prescribes an action at each step of the MAS. There are two main classes of strategies: \emph{memoryless} and \emph{memoryfull}. In the former case, agents choose an action by considering only the current state, while in the latter case, agents choose an action by considering the full history of the MAS.
Regarding information, we distinguish between \emph{perfect} and \emph{imperfect} information MAS. The former corresponds to a basic setting in which every agent has full knowledge about the MAS. However, real-life scenarios often involve situations where agents must act without having all relevant information at hand. In computer science, these situations occur, for example, when some variables of a system are internal/private and not visible to an external environment. In MAS models, imperfect information is usually modeled by setting an indistinguishability relation over the states of the MAS. This feature deeply impacts on the MC complexity. For example,  ATL and SL become undecidable in the context of imperfect information and memoryfull strategies~\cite{DT11}.

In this paper, we present a review of results on model checking for multi-agent systems concerning strategy classes and agent visibility, and explore potential directions for future research.

\paragraph{Outline} In Section~\ref{models}, we recall the definition of concurrent game structures and the syntax of ATL and SL. Then, in Section~\ref{strategies}, we review the main classes of strategies and the concept of information. In Section~\ref{complexity}, we revisit the model checking complexities for the aforementioned contexts. Finally, we conclude by providing some future directions in Section~\ref{conclusions}.

\section{Model and logics for MAS}\label{models}
In this section, we recall the definition of a formal model for MAS and the syntax of two well-known logics for strategic reasoning: ATL and SL.

We start by recalling the definition of concurrent game structures~\cite{AlurHK02}.
\begin{definition}\label{cgs}
	Given sets $Ag$ of agents and $AP$ of atoms, a \emph{concurrent game structure (CGS)} is a tuple 
	$M = \langle S, s_0, \{Act_i\}_{i \in Ag}, \allowbreak \{\sim_i\}_{i \in Ag}, d, \delta, V \rangle$ such that: 
	\begin{itemize} 
		\item $S$ is a finite, non-empty set of {\em states}, with {\em initial state}  $s_0 \in S$.
		
		\item For every $i \in Ag$, $Act_i$ is a finite, non-empty  set of {\em actions}. 
		
		Let $Act = \bigcup_{i \in Ag} Act_i$ be the set of all actions, and $ACT = \prod_{i \in Ag} Act_i$ the set of all {\em joint actions}, \textit{i.e.}, tuples of actions.
		
		\item For every $i \in Ag$, $\sim_i$ is a relation of {\em indistinguishability} between states, that is, an equivalence 
		relation on $S$.
		Given states $s, s' \in S$, $s \sim_i s'$ iff $s$ and $s'$ are said to be {\em observationally indistinguishable} for agent $i$. 
		
		\item The {\em protocol function} $d: Ag \times S \rightarrow (2^{Act}\setminus \emptyset)$ defines the availability of actions so that for every $i \in Ag$, $s \in S$, (i) $d(i,s) \subseteq Act_i$ and (ii) $s \sim_i s'$ implies $d(i,s) = d(i,s')$.
		
		\item The {\em transition function} $\delta : S \times ACT \to S$ assigns a successor state $s' = \delta(s, \vec{\alpha})$ to each state $s\in S$, for every joint action $\vec{\alpha} \in ACT$ such that $a_i \in d(i,s)$ for every $i \in Ag$, that is, $\vec{\alpha}$ is {\em enabled} at $s$. 
		 
		\item $V: S \times AP \rightarrow \{\top,\bot\}$ is a {\em two-valued labelling function}.
	\end{itemize}
\end{definition}

According to Def.~\ref{cgs}, a CGS describes the interactions of a group $Ag$ of agents, starting from the initial state $s_0 \in S$, following the transition function $\delta$. The latter is constrained by the availability of actions to agents, as specified by the protocol function $d$. 
A history $h \in S^+$ is a finite (non-empty) sequence of states.

Now, we recall the syntax of Alternating-time Temporal Logic~\cite{AlurHK02}.

\begin{definition}[$ATL^*$] \label{def:ATL*}
	The state ($\varphi$) and path ($\psi$) formulas in ATL$^*$ are defined as
	follows, where $p \in AP$ and $\Gamma \subseteq Ag$:
	\begin{eqnarray*}
		\varphi & ::= & p \mid \neg \varphi  \mid \varphi \land \varphi \mid \EATL{\Gamma} \psi\\
		\psi & ::= & \varphi \mid \neg \psi \mid \psi \land \psi \mid X \psi \mid (\psi U \psi) \mid (\varphi R \varphi)
	\end{eqnarray*}
	
	Formulas in ATL$^*$ are all and only the state formulas.
\end{definition}

ATL extends Computation Tree Logic (CTL)~\cite{ClarkeE81} in which the existential $E$ and
the universal $A$ path quantifiers are replaced with strategic modalities of the form $\EATL{\Gamma}$ and $\AATL{\Gamma}$,
where $\Gamma$ is a set of agents.
A formula $\EATL{\Gamma} \psi$ is read as \emph{``the agents in
coalition $\Gamma$ have a strategy to achieve $\psi$''}. The meaning of
temporal operators {\em next} $X$ and {\em until} $U$
is standard \cite{clarke1999model}.  Operators {\em unavoidable} $\AATL{\Gamma}$, {\em release} $R$, {\em finally} $F$, and
{\em globally} $G$ can be introduced as usual.

The formulas in the ATL fragment of ATL$^*$ are obtained from
Def.~\ref{def:ATL*} by restricting path formulas $\psi$ to the temporal operators.

To conclude this section, we recall the syntax of Strategy Logic~\cite{MogaveroMPV14}.

\begin{definition}[SL Syntax]\label{def:sl(syntax)}
	Given the set  $AP$ of atoms, variables $Var$, and agents $Ag$, the
	formal syntax of SL is defined as follows, where $p \in AP$, $x \in Var$, and $a \in Ag$:
	\begin{eqnarray*}
		\varphi &::=& p \mid \neg \varphi 
		\mid \varphi \wedge \varphi
		\mid \varphi \vee \varphi
		\mid X \varphi 
		\mid \varphi \:U \varphi
		\mid \varphi \,R\, \varphi
		\mid \ESL{x} \varphi 
		\mid \ASL{x} \varphi 
		\mid (a, x) \varphi 
	\end{eqnarray*}
\end{definition}

SL syntactically extends Linear-time Temporal Logic (LTL)~\cite{Pnueli77} with two \emph{strategy quantifiers}, the existential $\ESL{x}$ and universal $\ASL{x}$, along with an \emph{agent binding} $(a, x)$, where $a$ is an agent and $x$ a variable.  Intuitively, these additional elements can be respectively interpreted as \emph{``there exists a strategy $x$''}, \emph{``for all strategies $x$''}, and \emph{``bind agent $a$ to the strategy associated with the variable $x$''}.

\section{Classes of Strategies and Information}\label{strategies}
In this section, we recall some definitions of strategies and of agents' information.

In Definition~\ref{cgs}, we have defined an indistinguishability relation for each agent involved in the model. When every $\sim_i$ is the identity relation, \textit{i.e.}, $s \sim_i s'$ iff $s = s'$, we obtain a CGS with \emph{perfect information}~\cite{AlurHK02}.
When the latter is not true, we assume that every agent $i$ has \emph{imperfect information} about the exact state of the system. That is, in any state $s$, $i$ considers all states $s'$ that are indistinguishable for $i$ from $s$ to be epistemically possible~\cite{fagin2004reasoning}.
The indistinguishability relations are extended to histories in a
synchronous, pointwise way, \textit{i.e.}, histories $h, h' \in S^+$ are {\em indistinguishable} for agent $i \in Ag$, or $h \sim_i h'$, iff (i) $|h| = |h'|$ and (ii) for all $j \leq |h|$, $h_j \sim_i h'_j$.

Now, we have all the ingredients to present the different definitions of strategies. First, we start with a class of strategies in which the agents determine their actions by considering only the current state of the MAS.

\begin{definition}[Memoryless Strategy] \label{memoryless}
	A \emph{memoryless strategy} for agent $i \in Ag$ is a function $f_i
	: S \to Act_i$ such that for each state $s \in S$, (i)
	$f_i(s) \in d(i, s)$; and (ii) $s \sim_i s'$ implies $f_i(s) = f_i(s')$.  
\end{definition}

By Def.~\ref{memoryless}, any strategy for agent $i$ must return actions that are enabled for $i$ (i.e. condition (i)). Additionally, whenever two states are indistinguishable for $i$, the same action is returned (i.e. condition (ii)). This latter introduces the concept of \emph{uniformity}, where an agent can select a strategy that adheres to its visibility. Notice that, for the case of perfect information, condition (ii) is satisfied by any function $f_i : S \to Act_i$.

The notion of memoryless strategy is considered too weak for an agent. For this reason, the concept of memoryfull strategy has been introduced.

\begin{definition}[Memoryfull Strategy] \label{memoryfull}
	A \emph{memoryfull strategy} for agent $i \in Ag$ is a function $f_i: S^{+} \to Act_i$ such that for all histories $h,
	h' \in S^{+}$, (i) $f_i(h) \in d(i, last(h))$; and (ii)
	$h \sim_i h'$ implies $f_i(h) = f_i(h')$.  
\end{definition}

As for the memoryless case, memoryfull strategies must adhere to the protocol function and indistinguishability relation.

Between these two approaches, i.e. memoryless and memoryfull, several different classes of bounded strategies have been proposed, including works by {\AA}gotnes and Walther~\cite{AgotnesW09}, Brihaye et al.~\cite{BrihayeLLM09}, Vester~\cite{Vester13}, and Belardinelli et al.~\cite{BelardinelliLMY22}. In this work, we focus our attention to natural strategies~\cite{JamrogaMM19b}. The idea behind natural strategies is to adopt the view of bounded rationality, and look at “simple” strategies in specification of agents’ abilities. This notion has been introduced in both ATL and SL in the context of perfect~\cite{JamrogaMM17} and imperfect information~\cite{BelardinelliJMM22,JamrogaMM19a}.
Here, we focus on the definition provided in~\cite{JamrogaMM19a}.

A natural strategies is an \emph{ordered list of guarded actions}, i.e., sequences of pairs $(\phi_i,\alpha_i)$ such that:
	$\phi_i$ is a condition, and
	$\alpha_i$ is an action.
%
That is, a natural strategy is a rule-based representation in which the first rule whose condition holds in the current execution of the MAS is selected, and the corresponding action is executed.

With respect to the nature of the conditions, it is possible to define different classes of natural strategies. We start by recalling the notion of \emph{uniform natural memoryless strategy}.

\begin{definition}
	In an uniform natural memoryless strategy for any agent $a$, the conditions are defined over epistemic formulas as follows:
	\begin{center}
		$\psi ::= \top \mid K_a\varphi \mid  \neg \psi \mid  \psi \land \psi$
		
		$\varphi ::= p \mid  \neg \varphi \mid  \varphi \land \varphi \mid  K_b\varphi$
	\end{center}
	where $p$ is an atomic proposition, $b$ an agent, and $K_i$ is the knowledge operator for any agent $i$. Intuitively, a formula $K_i \varphi$ can be interpreted as \emph{``the agent $i$ knows $\varphi$''}.
\end{definition}

So, we have formulas that are prefixed by $K_a$ and then possibly combined by boolean operators. In other words, the formulas are always boolean conditions on $a$'s knowledge. 
As discussed in~\cite{JamrogaMM17}, to define \emph{natural memoryless strategy} in the perfect information case, we can replace epistemic formulas with boolean formulas only.

To improve the abilities of the agents, natural strategies have also been defined with recall. In particular, to evaluate properties over histories instead of states, a way to define conditions is to use regular expressions with the standard constructors $\cdot, \cup, *$ representing concatenation, nondeterministic choice, and finite iteration, respectively.
Thus, to define a \emph{natural strategies with recall} in the perfect information context, we can use regular expressions over boolean formulas. Similarly, to define a \emph{uniform natural strategies with recall} in the imperfect information context, we can use regular expressions over epistemic formulas. 

In the next section, we will provide the main model checking results for the above mentioned classes.

\section{Model Checking Complexities}\label{complexity}
Here, we discuss the model checking complexities for ATL and SL in terms of memoryless, memoryfull, and natural strategies in the perfect and imperfect information context.

First, we can analyze ATL. As you can see in Table~\ref{mcATL}, in the worst case, that is imperfect information and perfect recall strategies, the problem becomes undecidable, while in the perfect information case, the problem is polynomial. An interesting point of ATL is that memoryless and memoryfull strategies are equivalent in the perfect information case. This is because ATL is too weak in expressive power.

\begin{table}
\begin{center}  
	\begin{tabular}{|c|c|c|}
		\hline
		ATL & {\bf perfect information} & {\bf imperfect information}\\ 
		\hline 
		{\bf memoryless} & $PTIME$-complete~\cite{AlurHK02}
		& $\Delta_2^P$-complete~\cite{Schobbens04}
		\\
		\hline
		{\bf memoryfull} & $PTIME$-complete~\cite{AlurHK02}
		& undecidable~\cite{DT11}
		\\
		\hline
	\end{tabular}\caption{Model checking complexities for ATL.}\label{mcATL}
\end{center}
\end{table}

To overcome this problem, there is ATL$^*$. In this logic, strategic and temporal operators are decoupled to express more complex strategic objectives. As you can see in Table~\ref{mcATL*}, the model checking complexity becomes solvable in polynomial space in the memoryless case, double-exponential time in the memoryfull and perfect information case, and again undecidable in the memoryfull and imperfect information case. To address the latter problem, some works have either focused on an approximation to perfect information~\cite{BelardinelliFM23}, developed notions of bounded memory~\cite{BelardinelliLMY22}, or employed hybrid techniques~\cite{FerrandoM22,FerrandoM23}. Despite its expressiveness, ATL$^*$ suffers from the strong limitation that strategies are treated only implicitly in the semantics of strategic operators. This restriction
makes the logic less suited to formalize several important solution concepts, such as the Nash Equilibrium~\cite{myerson1991game}. 

\begin{table}
	\begin{center}  
		\begin{tabular}{|c|c|c|}
			\hline
			ATL$^*$ & {\bf perfect information} & {\bf imperfect information}\\ 
			\hline 
			{\bf memoryless} & $PSPACE$-complete~\cite{Schobbens04}
			& $PSPACE$-complete~\cite{Schobbens04}
			\\
			\hline
			{\bf memoryfull} & $2EXPTIME$-complete~\cite{AlurHK02}
			& undecidable~\cite{DT11}
			\\
			\hline
		\end{tabular}
	\end{center}\caption{Model checking complexities for ATL$^*$.}\label{mcATL*}
\end{table}

To gain expressive power, we need to move to Strategy Logic. As shown in Table~\ref{mcSL}, the model checking problem becomes intractable in the memoryfull case. Given the relevance of this logic, several fragments have been proposed~\cite{MogaveroMPV14,BelardinelliJKM19}. Among others, we would like to mention Strategic Logic One Goal that has the same model checking complexity as ATL$^*$ but more expressive power, and Strategic Logic Simple Goal that has the same model checking complexity as ATL but more expressive power.

\begin{table}
	\begin{center}  
		\begin{tabular}{|c|c|c|}
			\hline
			SL & {\bf perfect information} & {\bf imperfect information}\\ 
			\hline 
			{\bf memoryless} & -
			& $PSPACE$-complete~\cite{MaubertMMP21}
			\\
			\hline
			{\bf memoryfull} & non-elementary~\cite{MogaveroMPV14}
			& undecidable~\cite{DT11}
			\\
			\hline
		\end{tabular}
	\end{center}\caption{Model checking complexities for SL.}\label{mcSL}
\end{table}

\begin{table}
	\begin{center}  
		\begin{tabular}{|c|c|c|}
			\hline
			NatATL & {\bf perfect information} & {\bf imperfect information}\\ 
			\hline 
			{\bf memoryless} & $\Delta_2^P$-complete~\cite{JamrogaMM19b}
			& $\Delta_2^P$-complete~\cite{JamrogaMM19a}
			\\
			\hline
			{\bf with recall} & $PSPACE$-complete~\cite{JamrogaMM19b}
			& $PSPACE$-complete~\cite{JamrogaMM19a}
			\\
			\hline
		\end{tabular}
	\end{center}\caption{Model checking complexities for NatATL.}\label{mcNatATL}
\end{table}

In all the above-mentioned logics, the model checking problem is undecidable in the worst case. In the last few years, a natural way to represent strategies has been studied. From the definition of natural strategies, two variants of ATL and SL have been proposed. In these variants, called NatATL~\cite{JamrogaMM19b} and NatSL~\cite{BelardinelliJMM22}, the strategic operators are equipped with graded modalities that represent the complexity of the natural strategies in achieving the temporal objectives. As you can see in Tables~\ref{mcNatATL} and~\ref{mcNatSL} the model checking problem has a complexity less than or equal $PSPACE$. 

\begin{table}
	\begin{center}  
		\begin{tabular}{|c|c|c|}
			\hline
			NatSL & {\bf perfect information} & {\bf imperfect information}\\ 
			\hline 
			{\bf memoryless} & - 
			& $PSPACE$-complete~\cite{BelardinelliJMM22}
			\\
			\hline
			{\bf with recall} & -
			& $PSPACE$-complete~\cite{BelardinelliJMM22}
			\\
			\hline
		\end{tabular}
	\end{center}\caption{Model checking complexities for NatSL.}\label{mcNatSL}
\end{table}

Can this approach solve all the problems related to model checking for MAS? Unfortunately (or fortunately for researchers), there are several open problems related to what we have summarized in this work. We will discuss some of them in the following section.

\section{Future Directions}\label{conclusions}
As promised throughout the paper, in this section, we will discuss some directions for future research. We will summarize these aspects with respect to the three main features related to the formal verification of multi-agent systems, namely: strategies, information, and logics.

\paragraph{About strategies: find the good representation.} As we briefly discussed along the paper, both memoryless and memoryful strategies are not suitable choices. In fact, memoryless and memoryfull are too weak and strong, respectively, to describe agents' abilities. The weakness of memoryless strategies is a gain in terms of complexity, and the strength of memoryfull strategies is paid in terms of complexity. In addition, memoryfull strategies cannot be used to implement a model checker due to their domain over histories.
As we mentioned earlier, there are some bounded versions between memoryless and memoryfull, but there is a lot of work to do to standardize the good choice. For instance, in the context of natural strategies, there are various aspects that require attention, such as defining the appropriate notion of complexity for these strategies. In the above-mentioned works, the authors have proposed the total size of the strategy as complexity, i.e., the overall complexity is the sum of all the symbols involved in the conditions. Is this the best way to define the complexity of a strategy? This is an open problem that needs to be investigated. 
\vspace{-0.1em}
\paragraph{About information: perfect and imperfect is not enough.} We believe there is significant work to be done in this area. For instance, it seems too reductive to consider only white (i.e. perfect information) and black (i.e. imperfect information) settings. We advocate the need to define a taxonomy for imperfect information. As discussed earlier in the paper, some approaches try to make some MAS with imperfect information decidable. However, the authors in~\cite{BelardinelliJMM22,BelardinelliFM23,FM23} do not give a specific class of MAS with imperfect information that is decidable. Currently, only the class of hierarchical information has been proposed~\cite{Reif84,PnueliR90} and analyzed in SL~\cite{BerthonMMRV21}.
\vspace{-0.1em}
\paragraph{About logics: find the gap between ATL and SL.}
The two logics for the strategic reasoning discussed in this work suffer from two main problems on two different sides. On one hand, ATL has a good model checking complexity, but it cannot express several solution concepts such as the Nash Equilibrium. The strong limitation of ATL is that it treats strategies only implicitly in the semantics of its modalities. So, it is weak in expressiveness. On the other hand, SL is the more powerful logic for the strategic reasoning, but its model checking problem is not tractable. So, the full logic cannot have practical applications. The idea is to define a new logic for the strategic reasoning that can incorporate the positive features of ATL in terms of complexity and the good features of SL in terms of expressiveness. We understand that this is not a simple challenge, and finding a perfect trade-off between ATL and SL may be difficult. However, we see this need and want to go all the way on this point. 




\bibliographystyle{eptcs}

\end{document}